# Secure Quantum Communication and Superluminal Signalling on the Bell Channel

Remi Cornwall
Future Energy Research Group
Queen Mary, University of London, Mile End Road, London E1 4NS

## Abstract

A means and protocol is presented to send information on the Bell Channel to achieve the effect of superluminal signalling. The method is to use detection of a photon entangled state as one binary digit and either of the collapsed states as the complement digit – this is the protocol. The means to affect this detection is by use of an interferometer set-up able to resolve two interfering pathways corresponding to the two polarization states of the photon. To achieve interference of the horizontal and vertical components Faraday rotators are used to bring both components into diagonal polarization, this operation is unitary. Modulation is caused by the remote signaller collapsing one aspect of the photon wavefunction; a physically secure channel sending information superluminally results. A preliminary discussion into the clash and the hopeful resolution with Relativity theory is presented – it is noteworthy that at the instant of transmission between the two stations that there is no transfer of mass-energy to instigate communication but the transmission of a quantum state - *pure information* only.

Introduction

The formalism of Quantum Mechanics when dealing with a many bodied system requires a basis to span the variables of the system. Thus if we have an n-body system we could have a set of base states $|x_1..x_n\rangle$ for position, physical properties are derived from the wavefunction $|\psi\rangle$ on this basis. The state of the system evolves by a first order linear differential equation:

$$i\hbar \frac{d}{dt}|\psi\rangle = H|\psi\rangle \qquad \text{Eqn. 1}$$

This shows a totally deterministic evolution of the wavefunction, however measurement is not deterministic and the measurement M and $\langle\psi|M|\psi\rangle$ collapses into one of the eigenstates of the operator M. The EPR[1] paper asked if the formalism of QM was even correct by concocting a scenario of a two bodied system described by a wavefunction $\psi(x_1, x_2)$ in which the two particles were separated by a space-like interval and a measurement performed. It seemed that if the system was solely described by the wavefunction, a measurement of one of the particles would cause a 'collapse of the wavefunction' thus seeming to determine the physical property of the other distant particle *instantaneously*.

Einstein objected, wanting particles to have ascribed classical, objective properties and Special Relativity to be obeyed. Thus QM was seen as incomplete requiring hidden variables much as in a classical coin split down the middle and concealed in two black-boxes: one distant observer revealing 'heads' would know that the other distant observer had 'tails' the system already had a state that the measurement simply revealed. Other measurement paradoxes such as 'Schrödinger's Cat' highlighted deep philosophical problems too.

The way out of this quandary according to Bohr[2] and the principle of Complementarity (or Copenhagen Interpretation) was that one should not speak of unmeasured quantities as though they exist classically; we can only measure complementary pairs of observables that commute, thus $P_X$ and Y or $P_Y$ and X but not $P_X$ and X or $P_Y$ or Y. Aspects of measurement seem to complement each other and indeed place the system in the state permitted by the measurement. A glib rephrasing of this in a staunchly logical positivist frame is that nothing exists unless it is measured. Thus the EPR argument was misguided, in this viewpoint the measured values did not exist prior to measurement and there is no conspiracy to send information superluminally when the act of measurement and the *whole apparatus of measurement* is taken into account.

Meanwhile QM continued to have great successes and few were troubled by the apparent underlying philosophical non-objectivity. However some regarded Bohr's position as that of an obscurant and started to wonder if hidden variables existed and if this apparent superluminal communication was a real phenomena in rejection of the EPR view that it wasn't and could not be. Notably Bohm[3] (and de Broglie earlier) wondered if a 'quantum potential' or 'pilot wave' carrying only *information* could account for QM and place it back in a classical footing with addition of this device. Proofs were found that still required this hidden information to be sent superluminally and it was natural to wonder if it was real, something that could be tested experimentally. Bell[4,5] came up with a simplified EPR arrangement to test the predications of quantum over classical realism, the former causing correlations in the measurements over space-like intervals greater than the classical case. Figure 1 shows the essence of the setup where an entangled source of photons, S is incident on polarizing beam-splitters (PBS) and then detectors picking up the horizontal and vertical photons.

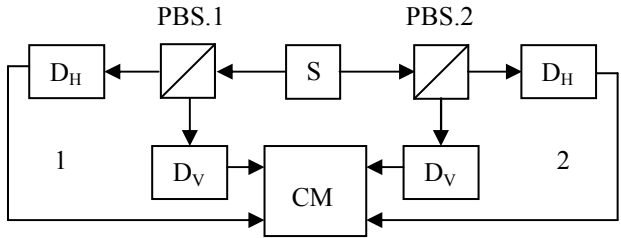

Figure 1 – Coincidence monitoring (CM) after detection ($D_X$) in an EPR type experiment

A coincidence monitor, CM can compute the expectation value of the signals at the detectors $D_H$ and $D_V$:

$$E(1, 2) = P_{HH}(1, 2) + P_{VV}(1, 2)$$
$$- P_{HV}(1, 2) - P_{VH}(1, 2)$$

The Bell inequality is computed, where the primes donate the PBSs at different angles:

$$| E(1,2) + E(1',2') + E(1',2) - E(1,2') | \leq 2$$
$$\text{Eqn. 3}$$

Noting the following probabilities:

$$P_{HH}(1, 2) = P_{VV}(1, 2) = \tfrac{1}{2}\cos^2(\theta_1 - \theta_2)$$
$$\text{and}$$
$$P_{HV}(1, 2) = P_{VH}(1, 2) = \tfrac{1}{2}\sin^2(\theta_1 - \theta_2)$$

Where $\theta_1$ is the angle of PBS1 and $\theta_2$ is the angle of PBS2. The expectation computes as:

$$E(1,2) = \cos2(\theta_1 - \theta_2)$$

and so forth for the other expectations.

For the so-called 'Bell Angles' of $\theta_1 = 3\pi/8$, $\theta_1' = 3\pi/8$ and $\theta_2 = \pi/4$, $\theta_2' = 0$ the Bell inequality is violated yielding:

$$| E(1,2) + E(1',2') + E(1',2) - E(1,2') | = 2\sqrt{2}$$

Alain Aspect[6] et al performed this and beyond most people's reasonable doubt it is known that *a posteriori* correlations could be discerned to have occurred between photon pair states on measurements. Newer experiments[7] over distances of up to 10km seem to make the space-like separation blunt. It is currently thought that signalling via this mechanism would be impossible from the indeterminacy of quantum measurement – modulation by a polarizer would result in our binary digit and its complement being signalled half of the time intended.

The Apparatus

Naively we cannot have the distant signaller collapse the wavefunction of an entangled photon into horizontal or vertical components and then have the distant receiver measure the complement to set up a scheme of binary communication. The act of measurement is indeterminate so if the signaller wants to collapse to a horizontal state, he will only achieve this half of the time – the signal becomes totally obfuscated in noise. Relativists still sceptical of the Bell Channel are delighted by this limit as it protects their sacrosanct mindset on causality and the scheme of things.

The indeterminacy of measurement can be overcome if we can use the non-collapsed state as a binary digit and either of the collapsed states as the other. Figure 2 shows a source (S) of entangled photons (pairs 1 and 2) as the communication channel. Distance between the polarising modulator and the interferometer is indicated by the double break in the lines showing the photon propagation. A non-destructive measurement[8,9] of the photon state by an interferometer set up (via polarising beam splitter, PBS) will distinguish the collapsed and non-collapsed states, it will distinguish the pure and mixed states as always regardless of the issue of entanglement. Note that at the instant of transmission photons are *already present* at the modulator and the detector - the signal is not transmitted by mass-energy only the quantum state is being transmitted.

Since the horizontal component will not interfere with the vertical component from the source both horizontal and vertical arms are rotated about the z-axis by a Faraday rotator or similar to bring them into diagonal alignment. To signal a binary 0 an entangled photon is sent via the communication channel (Table 1). This achieved by making the distant polarising filter transparent. At the interferometer the incident photons are set with a destructive interference length to give minimal signal. Binary 1 occurs when the filter is either horizontal or vertical such that un-entanglement is transmitted and maximum signal occurs at the detector because there is no destructive interference. Note that the interferometer is at a greater distance from the source than the modulator.

In reality several factors will make the probabilities deviate from the ideal: emission of un-entangled photons from the sources, imperfect optics and imperfect path lengths though it is an easy matter to amplify the difference between these two signals to achieve discrimination of the binary states.

Figure 2 – Transmitting Classical Data down a Quantum Channel

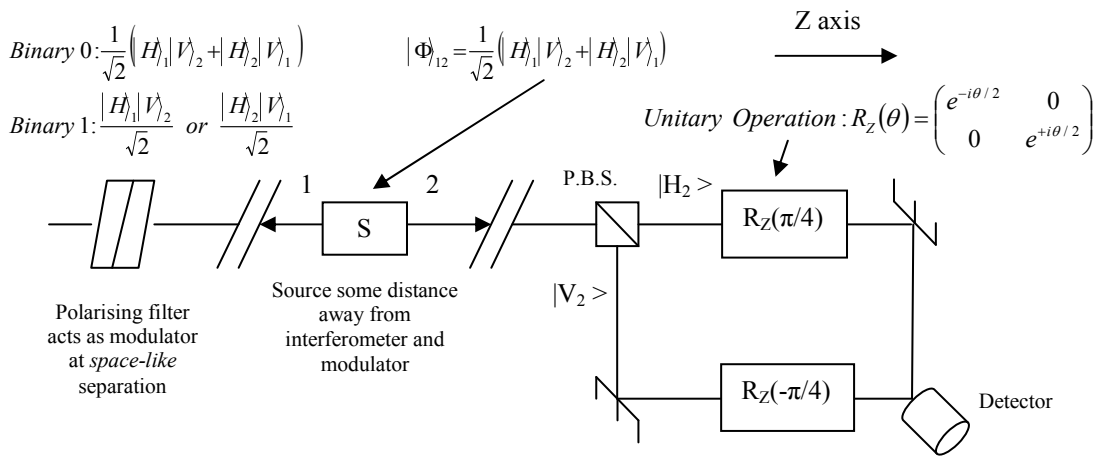

| Measurement/Modulation at distant system and state of two photon system | State of distant system | State of local system | Local measurement by _interferometer_ after modulation of distant system |
|---|---|---|---|
| No modulation: '**Binary 0**' $$\frac{1}{\sqrt{2}}\left(\left|H\right\rangle_1\left|V\right\rangle_2 + \left|H\right\rangle_2\left|V\right\rangle_1\right)$$ | Entangled => Pure state $$\frac{1}{\sqrt{2}}\left(\left|H\right\rangle_1 + \left|V\right\rangle_1\right)$$ (Or at least some superposition) | Entangled => Pure state $$\frac{1}{\sqrt{2}}\left(\left|V\right\rangle_2 + \left|H\right\rangle_2\right)$$ | Pure state results in interference (Or at least some interference since source is not ideally pure) |
| Modulation: '**Binary 1**' $$\frac{\left|H\right\rangle_1\left|V\right\rangle_2}{\sqrt{2}} \quad or \quad \frac{\left|H\right\rangle_2\left|V\right\rangle_1}{\sqrt{2}}$$ | Not entangled <=> Mixed state $$\frac{\left|H\right\rangle_1}{\sqrt{2}} \quad or \quad \frac{\left|V\right\rangle_1}{\sqrt{2}}$$ | Not entangled <=> Mixed state $$\frac{\left|V\right\rangle_2}{\sqrt{2}} \quad or \quad \frac{\left|H\right\rangle_2}{\sqrt{2}}$$ | Mixed state gives no interference |

Table 1 – The Protocol for Transmitting Classical Data down a Quantum Channel

It matters not that the beam is in an entirely pure state initially, at least some superposition and entanglement will result in maxima and minima at the detector when the beam is modulated and so render the protocol. The signal will 'ride on top' a large bias signal carrying no information but AC coupling from the detector to an amplifier can begin to discriminate. Several tens of photons are sent per bit to allow for path differences between the two arms of the interferometer and accurate interference but in principle one photon per bit is possible.

### A _Physically_ Secure Quantum Channel

Using two interferometers and modulators depicted in figure 2 a full duplex quantum channel can be set up. This channel is secure against "man in the middle attacks" because the information only exists at the extremities of the channel: any non-coherent measurement would collapse the wavefunction leaving only random noise; coherent measurement without the correct phase length would yield a constant binary digit as only entangled photons would be perceived. If the phase length could be guessed because the distance between the transmitting stations was well known, tapping into the channel would lead to massive obvious disruption and signal transmission loss; monitoring would catch this breach of security.

Nether-the-less further measures can be made by introducing a secret random phase length at both ends of the channel. The length of fibre optic cable, for instance, would be machine produced in matched pairs in a black box opaque to enquiry (by x-ray, ultrasound, terahertz radiation etc.) such that even the installer of the channel would not know the phase length. A security seal system too would destroy the apparatus if it was not inserted into the correct machinery of the communication channel but say time domain reflection equipment to ascertain the secret phase length. A secure docking procedure would do this.

A further aspect of the protection by the random phase length device would be if the eavesdropper was to guess a longer length as information exists after the modulation distance but not before. A periodic acknowledge-protocol within the permitted time frame of the channel phase length and the random phase length would ascertain that the wrong length has been inserted. Sub-nanosecond resolution would have the resolution to down to centimetres in a total channel length that could be kilometres. Phase lock would be a far from easy task.

Although the channel is quantum in nature, it is being used classically sending bits not qubits and all the conventional encryption measures for a classical digital channel would apply too. This physically secure and classically safe channel (in the sense of not cracking say, RSA codes should all the physical protection procedures be surmounted) is a boon to the transmission of sensitive information such as inter-bank money transfer or military information.

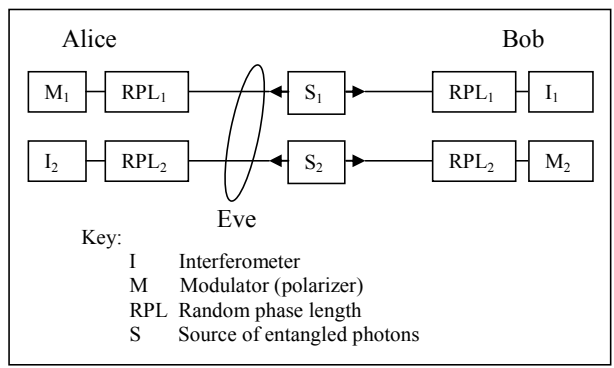

Figure 3 – A physically secure
quantum channel

### Discussion

An apparatus and argument has been presented for the instantaneous transmission of information as an adjunct to Bell's Theory and the Aspect experiments. Naturally there are concerns about conflicts with Relativity but it shall be shown that nature always must be sending information superluminally to ensure conservation of probability and a rational, consistent view of the universe emerges. Experiments exist already that show the effect of a 'quantum potential[3]' that carries only pure information such as repeated coherent interrogation/non-invasive measurement where the wavefunction feels out the experiment environment without transfer of energy to the object under investigation. Inescapably our view of space-time must be altered in the following presentation.

*Conservation of Probability Requires Superluminal Transfer of Quantum State Information*
The probability density of a normalised wavefunction in QM is given by the square of the wavefunction:

$$\rho(\mathbf{r},t) = |\psi(\mathbf{r},t)|^2$$
$$or$$
$$\int \rho(\mathbf{r},t) d^3 r = 1$$

If there is any sense in the concept, probability is conserved and would obey the continuity equation:

$$\frac{\partial \rho(\mathbf{r},t)}{\partial t} + \nabla \cdot \mathbf{j}(\mathbf{r},t) = 0$$

Where the probability current density $\mathbf{j}$ is derived on application of the Schrödinger equation to the above relations as:

$$\frac{\hbar}{2mi}\left(\psi^* \nabla \psi - \psi \nabla \psi^*\right)$$

Take a spherical source of particles (figure 4) emitted slowly enough to be counted one at a time. Arranged on a sphere one light-year in diameter (say) is a surface of detectors. Only one particle will be counted per detection event as the light-year diameter wavefunction collapses (becomes localised) randomly so that probability is conserved. The wavefunction, in current thought, is not perceived as something that is 'real' but is then discarded and a classical path is ascribed from the source to the detector that registered the event to say the particle, *retrospectively* went along that path.

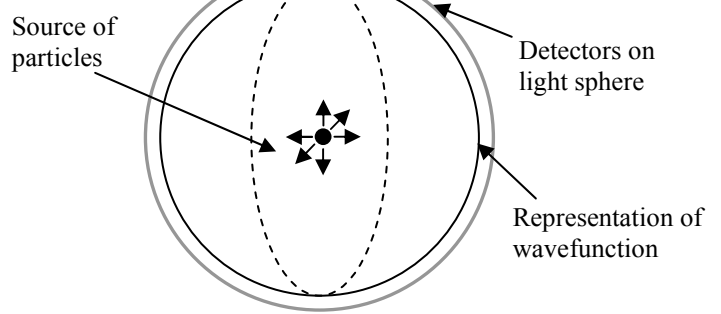

Source of particles

Detectors on light sphere

Representation of wavefunction

Figure 4 – Conservation of probability

There is however a problem of discarding the literality of the wavefunction and trying to apply classical concepts before measurement as exemplified by the delayed choice interference experiment (figure 5). Photons enter the apparatus incident on a half silvered mirror A. Two detectors 1 and 2 can elucidate what path the photon took as it came into the apparatus. A second half silvered mirror B inserted into the apparatus can cause the paths to interfere. If the interference length is set so that registry of a photon must mean that both arms of the interferometer were traversed, then this leads to a problem in the classical mode of though if once again we can expand the apparatus to gigantean proportions. Classically the photon (or particle) went along either arm but not both; the decision was made at mirror A. If the arms of our apparatus are light-years across, then inserting mirror B after the photon has entered the apparatus seems to be determining what path the photon went along or whether it decided to act as a wave and use both arms after it entered this apparatus.

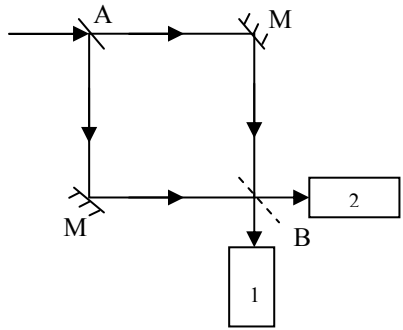

Figure 5 – Delayed choice interference experiment

Current thought, not really taking the truth of the wavefunction's *physical* existence gets into knots trying to explain these phenomena. We have seen the obfuscation of the Bohr/Copenhagen view where the photon doesn't really exist until it is measured - though something must have been travelling through space. The Many Worlds explanation needs a separate universe at each detection event scenario so that the Schrödinger equation is always obeyed at measurement. Another idea (working with one universe) is that the detector that registered the event sent information back to the first mirror to determine what path to take; this is the advanced and retarded wave formulation. The trouble here is with the delayed choice experiment - information went *back in time* in this viewpoint.

It is reasonable to apply Occam's Razor to interpretations of this quantum measurement process and admit in all simplicity, that nature is 'feeling' out the measurement environment across the whole of the wavefunction and is sending information superluminally. Thus in figure 4 the wavefunction interacts with the surface of detectors on the light sphere and *conspires* so that only one particle per event is recorded thus probability is conserved. Similarly in figure 5 the wavefunction traversed the apparatus and was incident on mirror B and the detectors to insure a consistent result. We suggest that nature has a scheme of keeping its state variables in check by superluminal transmission so concepts such as 'conservation of probability' aren't violated. The next section looks at interaction free measurement where an object can be imaged without, in the limit, photons being incident on it because it is interrogated by the *wavefunction*.

### Interaction Free Measurement by Repeated Coherent Interrogation

The picture that is being formed in this paper is the primacy of the wavefunction as a real object in physics and what the effect of its ability to communicate superluminally does to the current state of understanding of space-time in physics. The real world physical effects of the wavefunction cannot be questioned because of the field of quantum non-invasive measurement[8,9]. The essence of this is shown in the diagrams below:

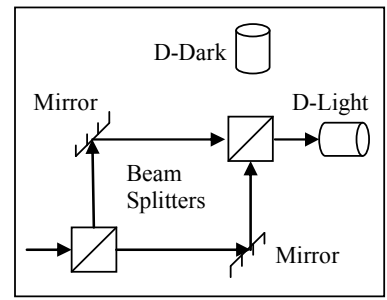

Figure 6a Interferometer with path length set for maximum signal at detector D-Light and minimum at D-Dark

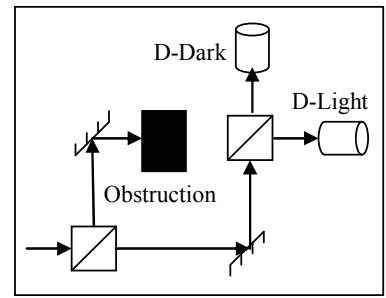

Figure 6b An obstruction destroys the interference at the second beam-splitter. By virtue of the signal at D-Dark we know 50% of the time that no photons interacted with the obstruction

Figure 6a shows an interferometer set up where a coherent photon source enters at the first beam splitter (partially silvered mirror) and recombines at a second. The detector D-Dark has its coherence length set so that the beams interfere destructively whilst the detector D-Light is set for constructive interference. In figure 6b an opaque object is placed in one arm of the interferometer. The firing of D-Dark indicates that a photon traversed the apparatus without interfering - that is it came down one arm only. Half of the time a photon will be absorbed by the object and the other half it will pass through to the detectors. We can say that the object has been detected with only half the incident number of photons into the measuring apparatus. Although beyond the scope of this paper figure 6c shows[8] the set up where by repeated coherent interrogations this 50% limit can be bettered and in the limit lead to no photons being absorbed by the object.

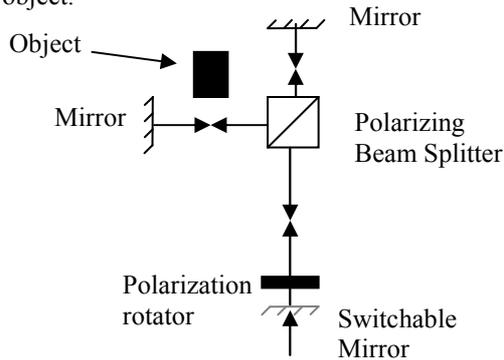

Figure 6c – Repeated coherent interrogation

The 'trick' here is that although the beam splitter, rotator and mirrors give a very low probability for the *photon* to enter the side arm with the object ($\delta$ is very small, $\sin^2 \delta \to 0$ in side arm, whilst main arm is $\cos^2 \delta \to 1$), the *wavefunction* always gets through, it is not attenuated (no potential barrier), we have $\psi = \sin \delta$ not say $\psi = A \sin \delta$ where A would be some attenuation factor. The wavefunction always measures the environment and can be made to traverse the apparatus many times *not* the photon. This gives a vanishing probability of photon interaction with the object but growing certainty of its presence. The lowest mirror switches out the interrogating *wavefunction* after a number of transits. A detector at a set interference length can work out if the side arm is blocked by the count of the detected *photons*.

*Simultaneity in Space, Simultaneity in Time*
The Lorentz Transform can be understood to have terms amounting to the transit time of light signals: Vt'$\gamma$ and Vx' $\gamma$ /c$^2$. The whole Lorentz group is then viewed as a rotation in the space-time of hyperbolic geometry. Absolute time and space concepts are gone; this is our view of 'reality'. What we say is that the physics is correct for light-speed signals (no change there!) but a better system of time measurement can be constructed with clocks using the Bell Channel. We suggest the transformation, x=x'$\gamma$ and t=t'$\gamma$ which can't be used to do physics (things respond to retarded potentials for instance) but is philosophically correct.

Below are two space-time diagram views of events very nearly simultaneous in time by a superluminal signal over a space-like interval with event A proceeding B. The Lorentz view gets causality wrong, whilst the 'expand and contract' view of the axis gets it right. Thus the quotidian (3 space + 1 time) view of <u>objective reality is restored to space</u>; events happen at a definite place and time agreeable by all observers – the Universe is a definite, objective stage in which the theatre of events occur. There is no need for an unknowable preferred reference frame in which simultaneity is preserved as Bell suggested – all observers can agree with this scheme and this was originally suggested by Lorentz in 1904 before reason was lost.

---

<u>The Lorentz transform</u>: $x = \gamma\left(x' + vt'\right)$  $t = \gamma\left(t' + \dfrac{vx'}{c^2}\right)$

Describes the transformation between inertial frames for different observers of mass-energy phenomena. All information about the co-ordinates is sent as mass-energy <u>too</u> so inevitably our measurement of space and time is affected (a bit like kicking a soccer ball whilst the goal posts are moving!).

This view point leads to the space-time construct, destruction of simultaneity in space and time (events A and B below) and the consideration of co-ordinate transformations as hyperbolic rotations in 4-space (hyperbolic 'angle' $\alpha$ in analogy to $\theta$ in 3-space rotations).

*Hyperbolic rotation matrix*

$u = (x_1 \ x_2 \ x_3 \ ict)$
$u' = L(\alpha)u$
$L = \begin{pmatrix} 1 & 0 & 0 & 0 \\ 0 & 1 & 0 & 0 \\ 0 & 0 & \cosh \alpha & i\sinh \alpha \\ 0 & 0 & -i\sinh \alpha & \cosh \alpha \end{pmatrix}$   *Where* $\alpha = \tanh^{-1}\dfrac{v}{c}$

Thus we obtain the familiar space-time diagram:

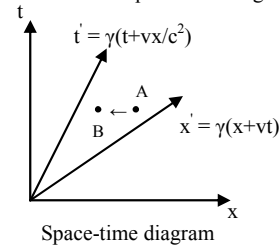

Space-time diagram

The terms in the Lorentz transform $\Delta x = \gamma v \Delta t'$ and $\Delta t = \gamma v \Delta x'/c^2$ can simply be understood as the <u>delay</u> in sending the information about the co-ordinates to the non-primed frame. For instance if it takes the primed frame $\Delta t'$ seconds to perform a measurement then the frame will have moved a distance $v\Delta t'$ which we correct back to the un-primed frame, $\gamma v\Delta t'$ in addition to any other distance measurement. As regards the time: the frame will have moved $v\Delta t'$ once again so the light signal will require an extra $v\Delta t'/c$ seconds to reach the source, now $\Delta t' = \Delta x'/c$ so the extra time is $\gamma v\Delta x'/c^2$ in the un-primed frame.

Sending information superluminally knocks out the terms $\Delta x = \gamma v\Delta t'$ and $\Delta t = \gamma v\Delta x'/c^2$ in the Lorentz transform giving the following transformation diagram:

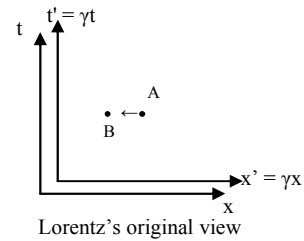

Lorentz's original view

---

Figure 7 – From space-time to Lorentz's view. Simultaneity of time and place is preserved. Note there is no reverse-causality associated with the Lorentz/SR transform and superluminal signals!

## Quantum Reality 1: Schrödinger's Equation in 3-Space

Superluminal effects and the physical existence of the wavefunction force us to change our view about space-time. What emerges is the primacy of movement in 3-space below the speed of light of the wavefunction with length and time dilation effects. The wavefunction carries information about a quantum particle through space to interact with other quantum systems such as the measuring device. We say something is a particle when it has been measured and regular concepts such as energy and momentum are ascribed to it. This classical intellectual baggage has us thinking in terms of particles moving through space when we really should be thinking in terms of the wavefunction as the primary concept. Operations on it such as $\psi^* E \psi$ define *physical observables* of the system from the *information* and hence the *physics*.

Indeed to bridge the gap between the classical and quantum worlds, textbooks ease our mind by showing us that in the classical limit where the action is large we get the geometric limit of particular paths and classical mechanics, thus the ray equation or the Hamilton-Jacobi Equations:

Solving the Schrödinger Equation for a single particle in three dimensions we obtain an approximation:

$$\psi = F e^{\frac{i}{\hbar} A} \qquad \text{Eqn. 4}$$

Where the phase A is a real function of co-ordinates that will be identified with the classical action and F is a real or complex function independent of time. Due to the smallness of h very rapid changes in phase result in this function over small distances; thus the wavefunction far away from the path of least action rapidly interferes and decays giving the notion of a classical path in the limit. Substitution of equation 4 in the Schrödinger Equation yields:

$$\left[ \frac{1}{2m} \left( \nabla^2 \mathbf{A} \right) + V + \frac{\partial \mathbf{A}}{\partial t} \right] F$$
$$- \frac{i\hbar}{2m} \left[ F \nabla^2 \mathbf{A} + 2 \left( \frac{\partial \mathbf{A}}{\partial x} \frac{\partial F}{\partial x} + \frac{\partial \mathbf{A}}{\partial y} \frac{\partial F}{\partial y} + \frac{\partial \mathbf{A}}{\partial z} \frac{\partial F}{\partial z} \right) \right]$$
$$- \frac{\hbar^2}{2m} \nabla^2 F = 0 \qquad \text{Eqn. 5}$$

By decreeing classical mechanics and letting h→0 which is equivalent to the wavelength going to zero, the 1$^{st}$ and 2$^{nd}$ order terms dropout yielding:

$$\frac{1}{2m} \left( \nabla^2 \mathbf{A} \right) + V + \frac{\partial \mathbf{A}}{\partial t} = 0 \qquad \text{Eqn. 6}$$

Which on the assumption that the wave is monochromatic and that:

$$\mathbf{A}(x, y, z, t) = \mathbf{S}(x, y, z) - h\nu t \Rightarrow S - Et$$

On substitution in equation 6 we obtain a form of the Hamilton-Jacobi Equation:

$$|grad \; \mathbf{S}| = \sqrt{2m(E - V)}$$

Somehow the quantum effects are wished out of view and we are further featherbedded by the idea of a particle in space being represented as a wave packet whose composition is given by the spectral Fourier coefficients. This applies when the particle has been measured and its position and momentum fall in a narrow range governed by the Uncertainty Principle such that a wave packet results. The situation in figure 4 invalidates this wave packet view point because the wavefunction is given by a spherical wave, $e^{i\mathbf{k}.\mathbf{r}}/r$ before measurement. It is only after detection that we ascribe position and momentum to a particle concept.

Really it is the wavefunction that travels through space, furthermore in figure 4 the wavefunction conspires with all the detectors such that conservation of probability is always true: if one photon is measured at one place at one time, it can be measured nowhere else. It is easier to apply Occam's razor to all the formulations of this measurement problem such as the Many Worlds, Advanced-Retarded Waves (the *pre-cognisance* of the measurement - even information travelling backwards in time from the future!) and admit in all simplicity that all the detectors have been superluminally connected by the wavefunction with passage of *information* such that only one photon per instant is measured.

It is convenient for the mind to show quantum mechanics as approximating classical mechanics. Via classical mechanics we derive our concepts of space and time, though we should stop trying to do this and face the quantum reality of the wavefunction moving through 3-space. Things exist at macroscopic level that can never be explained classically such as ferromagnetism, superconductivity, the shapes of molecules and the shapes of crystals and we should admit the same for space and time.

## Quantum Reality 2: The Measurement Problem and Decoherence

Quantum Mechanics is a description of nature and equation 1 should always be true. However measurement throws the system into an eigenstate of the measurement operator and assigns a probability to it thus:

$$state = \frac{M_M |\psi\rangle}{\sqrt{\langle \psi | M_M^* M_M | \psi \rangle}}$$

$$p(M) = \langle \psi | M_M^* M_M | \psi \rangle$$

This is the measurement problem: a non-unitary change from the Schrödinger equation to the above. Schrödinger highlighted this in his famous cat paradox where he showed a microscopic quantum event getting entangled with the macroscopic measurement equipment to magnify this obviously non-classical behaviour to absurd proportions. The result was that the cat was left in a superposition of the dead and alive states to be collapsed by when and by whom?

Some of the philosophical spin offs from this were Bohr's Complementarity/Copenhagen Interpretation, weird mind-body/consciousness effects collapsing the wavefunction, the Many World's Interpretation or advanced/retarded waves and quantum super-determinism in which events in the pre-ordained future affect the present. Applying Occam's Razor to this once again and noting what people are actually seeing in their attempts to construct quantum computers[10] and the difficulty of maintaining pure states, the most likely, sane candidate to explain the measurement problem is Decoherence Theory[11,12]. The central tenant of Decoherence Theory is the entanglement of a pure state with the environment and the calculation of the reduced density matrix for the system from the system-environment density matrix. Starting with a simple case, consider a closed two-state system described by the following state in two-dimensional Hilbert space (given pedagogically[12]):

$$|\psi\rangle = \alpha_0 |0\rangle + \alpha_1 |1\rangle$$

The states $|0\rangle$ and $|1\rangle$ are orthogonal. The most general way for calculating physical quantities in QM is by use of the density matrix/operator, thus:

$$\hat{\rho} = |\psi\rangle\langle\psi|$$

*giving*

$$\hat{\rho} = |\alpha_0|^2 |0\rangle\langle 0| + \alpha_0\alpha_1^* |0\rangle\langle 1| + \alpha_0^*\alpha_1 |1\rangle\langle 0| + |\alpha_1|^2 |1\rangle\langle 1|$$

Eqn. 7

*and the density matrix*

$$[\rho_{mn}] = [\langle m \mid \rho \mid n \rangle] = \begin{bmatrix} |\alpha_0|^2 & \alpha_0\alpha_1^* \\ \alpha_0^*\alpha_1 & |\alpha_1|^2 \end{bmatrix}$$

The diagonal components give the probability that the system is in either state, the off diagonal components the interference between the states. The expectation of any observable represented by an operator A is given by the trace over the product of the density and operator matrices:

$$\langle \psi \mid A \mid \psi \rangle = Tr(\rho A) = \sum_{mn} \rho_{mn} A_{mn}$$

The system cannot exist in isolation and through *unitary evolution* becomes entangled with the environment represented by states $|e_0\rangle$ and $|e_1\rangle$ which are in general non-orthogonal. On taking the tensor product, the density matrix becomes:

$$\hat{\rho}(t) = |\alpha_0|^2 |0\rangle \otimes |e_0\rangle\langle 0| \otimes \langle e_0| \quad + \quad \alpha_0\alpha_1^* |0\rangle \otimes |e_0\rangle\langle 1| \otimes \langle e_1|$$
$$+ \alpha_0^*\alpha_1 |1\rangle \otimes |e_1\rangle\langle 0| \otimes \langle e_0| \quad + \quad |\alpha_1|^2 |1\rangle \otimes |e_1\rangle\langle 1| \otimes \langle e_1|$$

In principle we cannot know the state of the environment and so we are left taking the *reduced density matrix* with the environmental states traced out. Orthogonal environment basis vectors $|e_0\rangle$ and $|e_0\rangle$ are used thus:

$$\langle e_0^\perp \mid e_0 \rangle = 0, \ \langle e_0 \mid e_1 \rangle = \cos\theta, \ \langle e_0^\perp \mid e_0 \rangle = \sin\theta$$

The reduced density matrix of the two-state system is given by:

$$\hat{\rho}_S(t) = Tr_E \rho(t) = \langle e_0 \mid \rho(t) \mid e_0 \rangle + \langle e_0^\perp \mid \rho(t) \mid e_0^\perp \rangle$$

*hence*

$$\hat{\rho}(t) = |\alpha_0|^2 |0\rangle\langle 0| + \alpha_0\alpha_1^* \cos\theta |0\rangle\langle 1|$$
$$+ \alpha_0^*\alpha_1 \cos\theta |1\rangle\langle 0| + |\alpha_1|^2 |1\rangle\langle 1| \quad \text{Eqn. 8}$$

Comparing this with eqn. 7 we see the modification to the coherence terms. The environmental states $e_0$ and $e_1$ are themselves evolving with time and since the environment is truly vast with many energy states, $e_0$ and $e_1$ will find themselves orthogonal in a very short period of time[11], for instance if each state is a function of many variables such as $(\mathbf{k}_1 \ldots \mathbf{k}_N, \mathbf{r}_1 \ldots \mathbf{r}_N)$ a change in *at least* one would lead to a very different wavefunction. Consider this simple example for part of the environment modelled by two particles in a rectangular box of infinite potential, the wavefunction for one particle is:

$$\psi_{n_1 n_2 n_3} = \sqrt{\frac{8}{abc}} \sin\frac{\pi n_1}{a} x . \sin\frac{\pi n_2}{b} y . \sin\frac{\pi n_3}{c} z$$

The dimensions of the box are a,b,c and taking the orthogonality condition for the two particles 1,2:

$$\int_V \psi_1 \psi_2 \, dxdydz = \delta_{abc}$$

Soon the wavefunctions are orthogonal - lattice vibrations/thermal relaxation effects will make a,b,c vary continuously in time.

Thus after a short time our environmental states become orthogonal and our density matrix tends to:

$$\hat{\rho}(t) = |\alpha_0|^2 |0\rangle\langle 0| + |\alpha_1|^2 |1\rangle\langle 1|$$

That is, a statistical mixture of pure states with no superposition. The whole density matrix evolves in a unitary manner but it is the act of taking the reduced trace, to that which concerns our system that gives the illusion of wavefunction collapse and non-unitary change. By the time we open the box, Schrödinger's Cat is already dead or still alive. A large statistical sample of such experiments would give the results of the reduced density matrix. We can't say which cat will live or die but only predict statistics exactly analogously to the probability space of a multi-particle problem in classical statistical mechanics.

## Conclusion

We have discussed a superluminal communication/encryption scheme. The 'Quantum Potential'[3] though pure information and having no mass-energy is real and engineering uses for it ought to be considered. It seems another trick has been squeezed out of nature similar to the amazement a century ago that greeted the Maxwell, Hertz, Marconi and Logie Baird discoveries of sending information, speech and pictures incredibly fast around the globe. Zeilinger et al[8,9] have talked about non-invasive measurements where X-rays could be used to image a source without actually (in the limit) imparting energy to the object – a boon to medical imaging perhaps. Understanding encryption, preserving it and working with it are crucial too for the burgeoning field of Quantum Computation[10].

At a fundamental level the process of entanglement of a quantum state with the environment seems to be giving some measure of understanding to this mysterious process and a semi-classical view of quantum mechanics becomes apparent with the wavefunction evolving deterministically by the Schrödinger Equation, always, as it should.

There is considerable irony here; Einstein disliked Quantum Mechanics for its apparent disregard for Objective Reality (indeterminacy and the measurement problem). Modern formulations of QM view the measurement problem as one of loss of coherence as a quantum system gets entangled with its environment[11]. This is a deterministic process as is the evolution of the isolated wavefunction anyway. Space-time with its denial of place and time really makes the universe a mystery, non-objective and *non-classical* – just how can we talk of the independent existence of an event if it is dependent on the measurement? The pot is calling the kettle black. Space-time is just a calculation/conceptualisation tool for effects involving mass-energy moving at or below the speed of light. Quantum Mechanics saves reason and returns the Universe to an <u>objective</u> stage of 3-space and time where simultaneous events and material things too can be said to have occurred or existed at a definite place and time independent of measurement. Classical 'sentiments' and intuition can return to physics in this way if we accept the primacy of a flow of the <u>quantum state</u> (and all that entails - the quantum rules) as a wave through 3-space and time (with relativistic effects of length contraction and time dilation) *instead* of a classical particle.

For Chris, Eugene and Farooq.